\documentclass[journal]{IEEEtran}

\usepackage{graphicx}
\usepackage{subcaption}
\usepackage{setspace}

\usepackage{color,soul}
\definecolor{lightblue}{rgb}{.60,.95,1}
\sethlcolor{lightblue}

\usepackage[cmex10]{amsmath}
\usepackage{amssymb}
\usepackage{amsthm}
\usepackage{MnSymbol}
\usepackage{float}

\usepackage{url}
\usepackage{cite}
\usepackage{bm}
\usepackage{multirow}
\usepackage{multicol}
\usepackage{enumerate}

\usepackage{algorithm}
\usepackage{algpseudocode}

\def\a{{\mathbf a}}
\def\n{{\mathbf n}}

\def\x{{\mathbf x}}
\def\y{{\mathbf y}}

\def\0{{\mathbf 0}}
\def\1{{\mathbf 1}}

\def\A{{\mathbf A}}

\def\I{{\mathbf I}}

\def\M{{\mathbf M}}

\def\S{{\mathbf S}}
\def\X{{\mathbf X}}
\def\Y{{\mathbf Y}}

\def\E{{\mathsf E}}

\def\bmu{{\boldsymbol{\mu}}}
\def\bbeta{{\boldsymbol{\eta}}}
\def\bgamma{{\boldsymbol{\gamma}}}

\def\bGamma{{\boldsymbol{\Gamma}}}

\def\bSigma{{\boldsymbol{\Sigma}}}

\def\cC{{\cal C}}
\def\cM{{\cal M}}
\def\cN{{\cal N}}

\def\bC{{\mathbb C}}

\def\trace{\text{Tr}}
\newcommand{\mtight}{\setlength{\thickmuskip}{0mu} \setlength{\medmuskip}{0mu} \setlength{\thinmuskip}{0mu}}

\begin{document}

\title{Sparse Bayesian learning with uncertainty models and multiple dictionaries}

\author{Santosh Nannuru, Kay L. Gemba, Peter Gerstoft, William S. Hodgkiss, Christoph Mecklenbr{\"a}uker%
\thanks{Santosh Nannuru, Kay L. Gemba, Peter Gerstoft, and William S. Hodgkiss are with the 
Scripps Institution of Oceanography at University of California, San Diego; snannuru@ucsd.edu,
gemba@ucsd.edu, pgerstoft@ucsd.edu, and whodgkiss@ucsd.edu.}%
\thanks{Christoph Mecklenbr{\"a}uker is with Institute of Telecommunications, Vienna University of Technology,
1040 Vienna, Austria, cfm@ieee.org.}
}

\maketitle

\begin{abstract}
Sparse Bayesian learning (SBL) has emerged as a fast and competitive method 
to perform sparse processing. The SBL algorithm, which is developed using a 
Bayesian framework, approximately solves a non-convex optimization problem 
using fixed point updates. It provides comparable performance and is significantly 
faster than convex optimization techniques used in sparse processing.
We propose a signal model which accounts for dictionary mismatch and the 
presence of errors in the weight vector at low signal-to-noise ratios. A fixed 
point update equation is derived which incorporates the statistics of mismatch 
and weight errors. We also process observations from multiple dictionaries.
Noise variances are estimated using stochastic maximum likelihood.
The derived update equations are studied quantitatively using beamforming 
simulations applied to direction-of-arrival (DoA). Performance of SBL using 
single- and multi-frequency observations, and in the presence of aliasing, is 
evaluated. SwellEx-96 experimental data demonstrates qualitatively the 
advantages of SBL.
\end{abstract}


\section{Introduction and motivation}
\label{sec:introduction}

Compressed sensing or sparse processing is the process of estimating sparse vectors 
using significantly fewer measurements. Mathematically, this corresponds to solving
an underdetermined system of linear equations under the constraint that the solution is sparse.
The exact solution has combinatorial complexity which is impractical to solve for high 
dimensional problems. The most popular, approximate and computationally feasible, 
sparse processing method is basis pursuit~\cite{chen2001} implemented using the 
LASSO~\cite{tibshirani1996} algorithm. Basis pursuit relaxes the sparsity criteria and 
the solution is given by solving a convex optimization problem. Though feasible, solving 
the optimization problem for high dimensions is still computationally slow. One of the faster 
alternatives is the matching pursuit algorithm~\cite{mallat1993}. But matching pursuit is a 
greedy approach and can lead to suboptimal support detection. Another alternative 
which is not greedy and is significantly faster than basis pursuit is sparse Bayesian 
learning (SBL)~\cite{tipping2001, wipf2004, wipf2007, ji2008, zhang2011, liu2012, gerstoft2016}.

In SBL, the sparse weight vector in the underdetermined system of linear equations is 
treated as a random vector with Gaussian prior. Explicit sparsity constraints are not
imposed on the weight vectors. Unlike traditional prior models, the parameters of the 
Gaussian prior are assumed unknown and are estimated by performing evidence 
maximization. The objective function for performing evidence 
maximization is non-convex and an approximate solution is obtained by formulating a 
fixed point update equation. The solution at convergence gives a parameter estimate 
which is sparse and hence the weight vectors are also sparse. 

A significant advantage of SBL over basis pursuit is that it can determine automatically 
the sparsity without any user input. Being a probabilistic approach, SBL computes the 
posterior distribution of the sparse weight vectors and hence provides estimates of their 
covariance along with the mean. Computationally, SBL can significantly outperform 
LASSO~\cite{gerstoft2016}.

Most of the literature on sparse processing assumes that the sensing matrix
or dictionary is deterministic and known. This is not feasible in many applications
such as beamforming~\cite{malioutov2005, xenaki2014} and matched-field 
processing~\cite{forero2014, gemba2016}. Also, at low signal-to-noise ratio
(SNR), the identified solution can contain false or spurious entries not present
in the true solution. These false entries often mask true entries and introduce 
errors in parameter estimation.

The three main contributions of this work are the following:
\newline
{\bf 1) SBL for uncertainty models:} We propose modifications to SBL to address sensing matrix 
mismatch and to reduce errors in the weight vector which occur in the presence of noise. 
The linear-Gaussian signal model is modified and transformed into a linear non-Gaussian model. 
Using approximations, the model remains linear-Gaussian and hence the regular SBL 
methodology can be applied. We focus on statistical modeling and integrating out of the error 
parameters rather than their estimation. This approach has the advantage that a large class of 
errors can be modeled and the resulting algorithm has a simple formulation. A portion of this
work addressing uncertainty in sensing matrix was published in~\cite{nannuru2017b}.
\newline
{\bf 2) Multi-snapshot and multi-dictionary SBL:} We derive an SBL algorithm for multiple 
snapshots using a fixed-point update~\cite{gerstoft2016}. This gives unbiased noise estimates 
and has better convergence properties especially for high SNR~\cite{gerstoft2016}. We then 
consider multi-dictionary observations with common sparsity profiles. When available, 
combining multi-dictionary observations using SBL provides a processing gain especially 
at low SNR as demonstrated with multi-frequency dictionaries~\cite{gemba2017}.
\newline
{ \bf 3) Simulations and real data analysis:} The proposed algorithms are demonstrated and 
verified using beamforming simulations for estimating direction-of-arrivals (DoAs) of multiple 
plane waves. Data from the SwellEx-96 experiment demonstrates application to real data 
and its ability to reduce aliasing when processing multiple frequencies.

The remainder of the paper is organized as follows. A brief literature review is provided in 
Sect.~\ref{sec:literature}. The signal model along with assumptions on priors and likelihoods 
are discussed in Sect.~\ref{sec:signal_model}. The SBL algorithm is derived in 
Sect.~\ref{sec:sbl} for uncertainty models and multiple dictionaries. The derived algorithms are 
studied using simulations and real data in Sect.~\ref{sec:simulations}. Conclusions are 
provided in Sect.~\ref{sec:conclusions}.

\subsection{Related literature}
\label{sec:literature}

SBL was introduced for regression and classification problems in the context of machine 
learning~\cite{tipping2001}. It has been used since for signal processing~\cite{wipf2004,
ji2008} with various modifications and extensions~\cite{wipf2007, zhang2011, liu2012}.

Since SBL does not impose explicitly any sparsity constraints but determines sparsity 
automatically, various explanations have been discussed. SBL solution can be obtained 
by solving an iterated reweighted LASSO problem and hence sparsity is 
expected~\cite{wipf2008, wipf2010}. Under certain conditions on the sensing matrix, SBL can 
identify sparse solutions without any explicit sparsity constraints~\cite{pal2014}. Cramer-Rao 
bounds for SBL solution are discussed in~\cite{prasad2013}. Various sparse signal recovery 
solutions including LASSO and SBL are unified within the Bayesian framework in~\cite{giri2016}.

Beamforming can estimate the DoAs of multiple plane waves from sensor
array observations. By formulating beamforming as an underdetermined linear problem, 
compressed sensing can estimate DoAs~\cite{malioutov2005, xenaki2014, gerstoft2015}.
The problem of mismatch and robustness of traditional beamforming algorithms has been 
studied extensively~\cite{cox1987, vorobyov2003, shahbazpanahi2003, lorenz2005, zhang2016}.

Perturbations and mismatch also have been addressed in the compressed sensing
literature for basis pursuit~\cite{herman2010, zhu2011, forero2014}, matching 
pursuit~\cite{teke2013}, and approximate message passing~\cite{parker2014}.
For SBL, beamforming in the presence of array imperfections is addressed 
in~\cite{liu2013, wu2016}. Robustness of SBL to outliers in the image processing 
application is studied in~\cite{fedorov2017}.

\subsection{Notation}
Scalar quantities are denoted by lowercase letters. A bold lowercase letter denotes 
a vector and a bold uppercase letter denotes a matrix. A vector or matrix of all zeros 
is denoted by $\0$ where appropriate dimensions are assumed. An identity matrix of 
dimension $N \times N$ is denoted $\I_{N}$. The notation $\M^{H}$ denotes the 
Hermitian (conjugate transpose). The transpose operation is denoted $\M^{T}$. 
The field of complex numbers is denoted $\bC$.

\section{Signal model}
\label{sec:signal_model}
In this section, we discuss the signal model used in SBL and the assumptions 
made in this paper. Let $\y \in \bC^{N}$ be the complex signal which is
expressed as
\begin{align}
\y &= \A \x + \n, \label{eq:linear_model}
\end{align}
where the noise $\n \in \bC^{N}$ is zero mean circularly symmetric complex Gaussian 
with density $\cC\cN(\n;\0,\sigma^{2} \I_{N})$; $\A \in \bC ^{N \times M}$ is the sensing 
matrix; $\x \in \bC^{M}$ is the weight vector. In sparse problem formulations, $\x$ is 
assumed sparse with at most $K$ non-zero entries where $\mtight K \ll M$. Sparsity 
level $K$ is not required explicitly or modeled by SBL. The vector $\x$ acts as a 
selection operator identifying columns of $\A$ that best explain the signal $\y$. 
We assume $\A$ has the maximal column rank $N$.

\emph{Error in sensing matrix:}
Often $\A$ is assumed known. This does not hold when there is uncertainty in the 
model or parameters used to construct $\A$. For example, in plane wave beamforming 
entries of $\A$ depend on array positions and wave speed which may be uncertain or 
can change over time. To account for perturbations we express
\begin{align}
\A &= \A^{o} + \A^{e}, \label{eq:A_error_model}
\end{align}
where $\A^{o}$ is known and $\A^{e}$ is a random perturbation 
matrix~\cite{swindlehurst1992, herman2010, zhu2011, teke2013, forero2014, zhang2016}.
For beamforming, sensing matrix perturbations have been studied 
in~\cite{vorobyov2003, liu2013}. An example where multiplicative noise gives rise
to such perturbations in the sensing matrix is discussed in Appendix A.
Though the component $\A^{e}$ is random and unknown, its statistics
are known. The prior model for $\A^{e}$ is discussed in Sect.~\ref{sec:prior_models}.

\emph{Error in weights:}
We assume $\x$ consists of two components
\begin{align}
\x &= \x^{o} + \x^{e}, \label{eq:x_error_model}
\end{align}
where the first component $\x^{o}$ is sparse and the second component $\x^{e}$ 
may be sparse. The vector $\x^{o}$ consists of the true complex weights whereas 
$\x^{e}$ is composed of errors in $\x$ due to noise or modeling mismatch. 
Likely $\x^{e}$ is sparse but we cannot uniquely distinguish the support of $\x^{e}$ 
from that of $\x^{o}$. Also, the support of $\x^{e}$ might vary because the noise 
realization changes over time. To overcome this limitation we assume that the 
statistics of $\x^{e}$ are known without knowledge of its support. Here both $\x^{o}$ 
and $\x^{e}$ are random and their prior models are discussed in Sect.~\ref{sec:prior_models}.

\emph{Signal model with uncertainty:}
Including the perturbed quantities from~\eqref{eq:A_error_model} and~\eqref{eq:x_error_model}, 
the signal model~\eqref{eq:linear_model} is
\begin{align}
\y &= \A \x + \n = (\A^{o} + \A^{e})(\x^{o} + \x^{e}) + \n \\
&= \A^{o} \x^{o} + \A^{e} \x^{o} + \A^{o} \x^{e} + \A^{e} \x^{e} + \n, \label{eq:linear_model_full}
\end{align}
where the first and the last terms are the regular linear model in SBL. 
The terms $\A^{e} \x^{o}$, $\A^{o} \x^{e}$ and $\A^{e} \x^{e}$ are additional 
``noise'' terms. We develop our theory for the general case and assume 
$\x^{o}$, $\x^{e}$, $\A^{e}$, and $\n$ are mutually independent. Since the
simulations (Sect.~\ref{sec:simulations}) consider either $\A^{e} = \0$ or 
$\x^{e} = \0$, the independence assumption of $\x^{e}$ and $\A^{e}$ is 
not crucial.

\subsection{Prior models}
\label{sec:prior_models}

\emph{Prior model for $\x^{o}$}:
In SBL $\x^{o}$ is modeled as a zero mean circularly symmetric complex Gaussian 
with prior density $p(\x^{o}) = \cC \cN(\x^{o} ; \0 , \bGamma)$, where the unknown 
covariance matrix $\bGamma$ is assumed diagonal,
$\bGamma = \text{diag}(\bgamma)$, $\bgamma = [\gamma_{1}  \ldots \gamma_{M}]$. 
The covariance $\bGamma$ is estimated by SBL.

We assume the error terms $\x^{e}$ and $\A^{e}$ are stochastic and define statistics
over them. These statistics easily integrate all possible error realizations while
computing evidence and allows us to study their effect on average. An alternate 
approach could be to estimate $\x^{e}$ and $\A^{e}$ from the data. This would 
increase significantly the dimensionality of the problem and is not pursued here.

\emph{Prior model for $\x^{e}$}:
The term $\x^{e}$ was introduced to account for errors in $\x$. We model $\x^{e}$ to have 
zero mean and known diagonal covariance $\bGamma^{e} =  \text{diag}(\bgamma^{e})$.
It quantifies the prior knowledge of errors 
in $\x$. We can choose $\bgamma^{e}$ empirically based on the specific application.
The term $\x^{e}$ establishes a noise floor for $\x$ and helps in strengthening 
weaker weights (see Sect.~\ref{sec:simulations}). In this sense it is similar to 
the concept of stochastic resonance~\cite{blondeau2004, chen2007} where adding 
noise into a non-linear system improves its detection performance.

\emph{Prior model for $\A^{e}$}:
Let $p(\A^{e})$ be the density function of the error matrix $\A^{e} = [\a^{e}_{1} \ldots \a^{e}_{M}]$.
For computational tractability we assume that the $m$th column $\a_{m}^{e}$ has known
 covariance $\bSigma_{m}^{e}$. No assumption is 
made about the mean. Also, let the columns of $\A^{e}$ be statistically orthogonal. Hence
\begin{align}
\E(\a_{m}^{e} \a_{n}^{e H}) &=  \delta(m-n) \bSigma_{m}^{e}. \label{eq:A_error}
\end{align}

In~\cite{forero2014} the perturbation vectors $\a_{m}^{e}$ are assumed 
stochastic and an elastic net regression is formulated by averaging out 
the perturbations. The perturbations are assumed to be complex Gaussian 
random vectors in~\cite{zhang2016}. Parametric modeling of the perturbations 
$\a_{m}^{e}$ is considered in~\cite{liu2013} for plane wave beamforming. The 
parameters are estimated within the iterative framework of SBL but only specific 
perturbations are considered and cannot be generalized to include a broader 
class of errors.

\subsection{Approximate likelihood}
\label{sec:likelihood}

Combining all the ``noise" terms together as $\bbeta = \A^{e} \x^{o} + \A^{o} \x^{e} + \A^{e} \x^{e} + \n$
gives
\begin{align}
\y &= \A^{o} \x^{o} + \bbeta. \label{eq:linear_model_new}
\end{align}
The modified noise $\bbeta$ is not Gaussian since $\bbeta$ is composed of terms
$\A^{e}$ and $\x^{e}$ whose densities are not known in general (from the prior models 
in Sect.~\ref{sec:prior_models}). To move forward within the SBL framework, we 
approximate $\bbeta$ to be Gaussian. Note that a Gaussian assumption on the
variables $\A^{e}$ and $\x^{e}$ still will not simplify the distribution of $\bbeta$ as the
terms $\A^{e} \x^{o}$ and $\A^{e} \x^{e}$ involve products of Gaussian random 
variables which do not have closed form distributions.

To simplify the likelihood model, we compute the mean and covariance of $\bbeta$: 
\begin{align}
\E(\bbeta) &= \E(\A^{e} \x^{o} + \A^{o} \x^{e} + \A^{e} \x^{e} + \n) = \0 \\
\bSigma_{\bbeta} &= \E(\bbeta \bbeta^{H}) = \E(\A^{e} \x^{o} \, \x^{o H} \A^{e H}) + \E(\A^{o} \x^{e} \, \x^{e H} \A^{o H}) \nonumber \\ 
& \qquad + \E(\A^{e} \x^{e} \, \x^{e H} \A^{e H}) + \E(\n \n^{H}) \\
&= \sum_{m,n} \Big[ \E(x^{o}_{m} x^{o H}_{n}) \E(\a^{e}_{m} \a^{e H}_{n}) + 
	\E(x^{e}_{m} x^{e H}_{n}) \a^{o}_{m} \a^{o H}_{n} \nonumber \\
& \qquad + \E(x^{e}_{m} x^{e H}_{n}) \E(\a^{e}_{m} \a^{e H}_{n}) \Big] + \sigma^{2} \I_{N} \\
&= \sum_{m} \Big[ \gamma_{m} \bSigma_{m}^{e} + \gamma^{e}_{m} \a^{o}_{m} \a^{o H}_{m}
	+ \gamma^{e}_{m} \bSigma_{m}^{e} \Big] + \sigma^{2} \I_{N} \label{eq:full_model_covariance}
\end{align}
We have used the independence of $\x^{o}$, $\x^{e}$, $\A^{e}$, and $\n$ in 
the above simplification. While computing the covariance of $\bbeta$, the error terms
$\x^{e}$ and $\A^{e}$ are integrated out and the covariance matrix $\bSigma_{\bbeta}$ 
depends on their statistics $\bgamma^{e}, \bSigma_{m}^{e}$ along with $\bgamma$
and $\sigma^{2}$. This integration circumvents the need to estimate explicitly the
unknowns $\x^{e}$ and $\A^{e}$.

For analytical simplification, we approximate the density of $\bbeta$ to be Gaussian
with mean zero and covariance $\bSigma_{\bbeta}$
\begin{align}
p(\bbeta) & \approx \cC \cN (\bbeta; \0, \bSigma_{\bbeta}). \label{eq:modified_noise}
\end{align}
To justify this approximation expand the modified noise as:
$\bbeta = \displaystyle \sum_{m}{(x^{o}_{m} \a^{e}_{m} + x^{e}_{m} \a^{o}_{m} + x^{e}_{m} \a^{e}_{m})} + \n$.
Thus $\bbeta$ is a sum of a large number of random vectors. From the central limit 
theorem, $\bbeta$ converges to a Gaussian distribution as 
$\mtight M \rightarrow \infty$. When $\x^o$ is $K$-sparse, the error in the Gaussian 
approximation \eqref{eq:modified_noise} decreases with $\frac{1}{\sqrt{K}}$.
The likelihood for the signal model~\eqref{eq:linear_model_new} is approximately
\begin{align}
p(\y | \x^{o}) &= p(\y | \x^{o} ; \A^{o}) \approx \cC \cN(\y ; \A^{o} \x^{o}, \bSigma_{\bbeta}). \label{eq:likelihood_new}
\end{align}
Once the modified noise $\bbeta$ is approximated as Gaussian, we treat $\bbeta$ 
and $\x^{o}$ as independent (which is not necessarily true from the expression for 
$\bbeta$). This assumption is necessary to evaluate analytically the evidence in
Sect.~\ref{sec:evidence}.

\subsection{Multiple snapshots}
To increase the SNR, we process multiple observations (snapshots) simultaneously. 
Let $\Y  = [ \y_{1} \ldots \y_{L} ] \in \bC^{N \times L}$ denote $L$ consecutive snapshots 
arranged column-wise in a matrix. The multi snapshot analogue of~\eqref{eq:linear_model} is
\begin{align}
\Y &= \A^{o} \X^{o} + \underline{\bbeta}
\end{align}
where $\X^{o} = [ \x_{1}^{o} \ldots \x_{L}^{o} ]$ 
and $\underline{\bbeta} = [ \bbeta_{1} \ldots \bbeta_{L} ]$. The $\x_{l}^{o}$ are assumed i.i.d. 
Gaussian across snapshots
\begin{align}
p(\X^{o}) &= \prod_{l=1}^{L}{p(\x_{l}^{o})} = \prod_{l=1}^{L}{\cC \cN(\x_{l}^{o} ; \0 , \bGamma)}.
\end{align}
The error terms $\A^{e}$, $\x^{e}$, and the noise $\n$ are assumed independent 
across snapshots. The multi-snapshot likelihood is 
\begin{align}
p(\Y | \X^{o}) &= \prod_{l=1}^{L}{p(\y_{l} | \x_{l}^{o}; \A^{o})}
\label{eq:ms_likelihood}
\end{align}
where the single snapshot likelihood $p(\y_{l} | \x_{l}^{o}; \A^{o})$ is in~\eqref{eq:likelihood_new}.

\subsection{Multiple dictionaries}
We assume observations generated by a set of dictionaries are available 
simultaneously and a portion of the support is common for all the weights.
We are interested in recovering this shared sparsity structure. A physical example 
are recorded observations at several frequencies but generated by the same 
sparse set of sources (see Sect.~\ref{sec:experimental_data}).

Let the observation vectors recorded by $F$ dictionaries be 
$\Y_{1:F} \equiv \{\Y_{1} \ldots \Y_{F}\}$ with the corresponding sparse weights 
$\X_{1:F}^{o} \equiv \{\X_{1}^{o} \ldots \X_{F}^{o}\}$. We have
\begin{align}
\Y_{f} &= \A_{f}^{o} \X_{f}^{o} + \underline{\bbeta}_{f} \, , \quad f = 1,\ldots,F
\end{align}
where $\A_{f}^{o}$ are the sensing matrices and $\underline{\bbeta}_{f}$ are (modified) noise 
contributions. The noise $\underline{\bbeta}_{f}$ and the weights $\X_{f}^{o}$ are assumed 
independent. The multi-dictionary likelihood is then
\begin{align}
p(\Y_{1:F} | \X_{1:F}^{o}) &= \prod_{f=1}^{F}{p(\Y_{f} | \X_{f}^{o})} \label{eq:MF_likelihood}
\end{align}
where $p(\Y_{f} | \X_{f}^{o})$ is given by~\eqref{eq:ms_likelihood}. We have two possibilities 
for the joint multi-dictionary prior over $\X_{1:F}^{o}$.

\emph{Multiple covariance (MC) prior:}
In this model, the joint prior is given by
\begin{align}
p(\X_{f}^{o}) &= \prod_{l=1}^{L}{\cC \cN(\x_{f,l}^{o} ; \0,\bGamma_{f})} \, , \quad f = 1,2,\ldots,F
\label{eq:MF_prior_1}
\end{align}
where the prior covariance $\bGamma_{f} = \text{diag}(\bgamma_{f})$ depends on the dictionary.
This model has been used in the context of multi-frequency beamforming in~\cite{gerstoft2016a}.

\emph{Common covariance (CC) prior:}
This model assumes the prior for all dictionaries is governed by the same statistical distribution
\begin{align}
p(\X_{f}^{o}) &= \prod_{l=1}^{L}{\cC \cN(\x_{f,l}^{o} ; \0,\bGamma)} \, , \quad f = 1,2,\ldots,F 
\label{eq:MF_prior_2}
\end{align}
i.e. $\bGamma_{1}  = \cdots = \bGamma_{F} = \text{diag}(\bgamma)$. This imposes 
identical sparsity constraints on $\X_{1}^{o} \ldots \X_{F}^{o}$. A common covariance 
matrix in multi-frequency beamforming was used in~\cite{liu2012}.

\section{Sparse Bayesian learning}
\label{sec:sbl}
\subsection{Evidence}
\label{sec:evidence}

In the SBL framework~\cite{tipping2001, wipf2007}, the prior parameter 
$\bgamma$ is assumed unknown and estimated using the observed signal 
$\Y$. It is estimated by maximizing the evidence (also called Type-II 
maximum likelihood). We first consider the single dictionary case. The 
evidence $p(\Y)$ is obtained by averaging over all realizations of $\X^{o}$
\begin{align}
& p(\Y) = \int{p(\Y | \X^{o}) p(\X^{o}) d\X^{o}} \\
&= \int{\prod_{l=1}^{L}{\cC \cN(\y_{l} ; \A^{o} \x_{l}^{o} , \bSigma_{\bbeta}) \, \cC \cN(\x_{l}^{o} ; \0 , \bGamma)} d\X^{o}} \\
&= \prod_{l=1}^{L}{\cC \cN(\y_{l} ; \0, \bSigma_{\bbeta} + \A^{o} \bGamma \A^{o H})} 
	= \prod_{l=1}^{L}{\cC \cN(\y_{l} ; \0, \bSigma_{\y})}, \nonumber
\end{align}
where $\bSigma_{\y} = \bSigma_{\bbeta} + \A^{o} \bGamma \A^{o H}$ and it depends 
on the parameters $\sigma^2$ and $\bgamma$. Ignoring the terms independent of 
$\sigma^2$ and $\bgamma$
\begin{align}
\log p(\Y) &= \sum_{l=1}^{L}{ - \log \big( \pi^{N} |\bSigma_{\y}| \big)} - \sum_{l=1}^{L}{\y_{l}^{H} \bSigma_{\y}^{-1} \y_{l}} \\
& \propto - L \log |\bSigma_{\y}| - \trace (\Y^{H} \bSigma_{\y}^{-1} \Y),
\end{align}
where $\trace()$ denotes the trace of a matrix.

\subsection{Fixed point update}
\label{sec:FPU_regular}

The estimate $\hat{\bgamma}$ maximizes the evidence
\begin{align}
\hat{\bgamma} &= \underset{\bgamma}{\arg \, \max} \, \log \, p(\Y) \\
&= \underset{\bgamma}{\arg \, \min} \,  \Bigg\{ L \log |\bSigma_{\y}| + 
	\trace (\Y^{H} \bSigma_{\y}^{-1} \Y) \Bigg\}. \label{eq:objective}
\end{align}
One approach to solve this problem is to use the EM algorithm~\cite{dempster1977}
but the resulting update equations have slow convergence~\cite{tipping2001, wipf2007}.
We perform differentiation of the objective function~\eqref{eq:objective} to obtain 
a local minimum. We have the following derivative relations for $\bSigma_{\y}$
\begin{align}
\frac{\partial \, \log |\bSigma_{\y}|}{\partial \gamma_{m}} 
&= \trace\Bigg(  \bSigma_{\y}^{-1} \frac{\partial \bSigma_{\y}}{\partial \gamma_{m}} \Bigg ), \\
\frac{\partial \, \bSigma_{\y}^{-1}}{\partial \gamma_{m}} 
	= - \bSigma_{\y}^{-1} \frac{\partial \bSigma_{\y}}{\partial \gamma_{m}} \bSigma_{\y}^{-1}, & \quad
\frac{\partial \bSigma_{\y}}{\partial \gamma_{m}} = \bSigma_{m}^{e} + \a_{m}^{o} \a_{m}^{o H}.
\end{align}
Differentiating~\eqref{eq:objective} with respect to the $m$th diagonal element 
$\gamma_{m}$
\begin{align}
\frac{\partial}{\partial \gamma_{m}} & \Bigg\{ L \log |\bSigma_{\y}| + \trace (\Y^{H} \bSigma_{\y}^{-1} \Y) \Bigg\} \nonumber \\
&= L \, \trace\Big(  \bSigma_{\y}^{-1} [\bSigma_{m}^{e} + \a_{m}^{o} \a_{m}^{o H}] \Big ) - \nonumber \\
& \qquad \quad \trace \Big( \Y^{H} \bSigma_{\y}^{-1} [\bSigma_{m}^{e} + \a_{m}^{o} \a_{m}^{o H}] \bSigma_{\y}^{-1} \Y \Big).
\end{align}
Equating the derivative of the objective function to zero 
\begin{align}
& 1 = \frac{1}{L} \frac{\trace \Big( \Y^{H} \bSigma_{\y}^{-1} [\bSigma_{m}^{e} + \a_{m}^{o} \a_{m}^{o H}] \bSigma_{\y}^{-1} \Y \Big)}
	{\trace \Big(  \bSigma_{\y}^{-1} [\bSigma_{m}^{e} + \a_{m}^{o} \a_{m}^{o H}] \Big)} \\
& \frac{\gamma_{m}}{\gamma_{m}} 
= \Bigg( \frac{1}{L} \frac{\trace \Big( \Y^{H} \bSigma_{\y}^{-1} [\bSigma_{m}^{e} + \a_{m}^{o} \a_{m}^{o H}] \bSigma_{\y}^{-1} \Y \Big)}
	{\trace \Big(  \bSigma_{\y}^{-1} [\bSigma_{m}^{e} + \a_{m}^{o} \a_{m}^{o H}] \Big)} \Bigg)^{b}
\end{align}
where we introduced $\gamma_{m}$ terms to obtain an iterative update equation.
Since the fixed point update is not unique, the exponent term $b$ is introduced to 
include a broad range of update rules. Different update equations introduced in the 
literature can be obtained using different values of $b$. The update then is
\begin{align}
\gamma_{m}^{\text{new}} = \gamma_{m}^{\text{old}}  
	\Bigg( \frac{\trace \Big( \bSigma_{\y}^{-1} [\bSigma_{m}^{e} + \a_{m}^{o} \a_{m}^{o H}] \bSigma_{\y}^{-1} \S_{\y} \Big)}
	{\trace \Big(  \bSigma_{\y}^{-1} [\bSigma_{m}^{e} + \a_{m}^{o} \a_{m}^{o H}] \Big)} \Bigg)^{b}. \label{eq:gamma_update}
\end{align}
where $\S_{\y}$ is the sample covariance matrix $\S_{\y} = \frac{1}{L} \Y \Y^{H}$.
The SBL update~\eqref{eq:gamma_update} incorporates statistics ($\bSigma_{m}^{e}$ and 
$\bgamma^{e}$) of uncertainty models.

\emph{Remark:} There are multiple ways to formulate a fixed point update equation.
Our formulation is inspired by some of the equations used in the 
literature~\cite{tipping2001, wipf2007, gerstoft2016} and convergence properties of the 
simulation results. It is not clear for what values of $b$, if any, convergence 
of~\eqref{eq:gamma_update} is guaranteed. For $\bSigma_{m}^{e} = \0$ 
and $\bgamma^{e} = \0$, a value of $b = 1$ gives the update equation used 
in~\cite{tipping2001, wipf2007} and $b = 0.5$ gives the update equation 
in~\cite{gerstoft2016}.

\subsection{Multi-dictionary SBL}
\label{sec:MF_SBL}
We have two multi-dictionary update rules based on the priors for $\X_{1:F}$
in either \eqref{eq:MF_prior_1} or \eqref{eq:MF_prior_2}.

\subsubsection{SBL-MC}
With the prior \eqref{eq:MF_prior_1} that is dictionary-dependent, the likelihood 
\eqref{eq:MF_likelihood}, and the independence assumptions, the joint evidence 
$p(\Y_{1:F})$ is
\begin{align}
p(\Y_{1:F}) &= \prod_{f=1}^{F}{p(\Y_{f})} = \prod_{f=1}^{F}{ \prod_{l=1}^{L}{\cC \cN(\y_{f,l} ; \0 ,\bSigma_{\y_{f}})}}.
\label{eq:joint_evidence}
\end{align}
where $\bSigma_{\y_{f}} = \bSigma_{\bbeta_{f}} + \A_{f}^{o} \bGamma_{f} \A_{f}^{o H}$.
Since the different dictionary components are decoupled, maximizing the joint evidence 
corresponds to maximizing the evidence for each dictionary individually. Thus the 
update rule for $f$th dictionary is 
\begin{align}
\gamma_{f,m}^{\text{new}} & = \gamma_{f,m}^{\text{old}}  
	\Bigg( \frac{\trace \Big( \bSigma_{\y_{f}}^{-1} [\bSigma_{f,m}^{e} + \a_{f,m}^{o} \a_{f,m}^{o H}] \bSigma_{\y_{f}}^{-1} \S_{\y_{f}} \Big)}
	{\trace \Big(  \bSigma_{\y_{f}}^{-1} [\bSigma_{f,m}^{e} + \a_{f,m}^{o} \a_{f,m}^{o H}] \Big)} \Bigg)^{b}. \label{eq:gamma_update_MF2a}
\end{align}
We can combine $\bgamma_{f}$ to obtain a multi-dictionary estimate 
\begin{align}
\bgamma &= \frac{1}{F} \sum_{f=1}^{F} \bgamma_{f}. \label{eq:gamma_update_MF2b}
\end{align}
If the sparsity of $\bgamma_{f}$ is the same across dictionaries, the averaging above will 
enhance the sparsity of the estimate $\bgamma$ in presence of noise. The summation 
\eqref{eq:gamma_update_MF2b} is inspired by traditional multi-frequency processing in
conventional beamforming where the beamformer outputs at each frequency are combined
incoherently~\cite{gemba2017}.

\subsubsection{SBL-CC}
With the prior~\eqref{eq:MF_prior_2} that is common across dictionaries, the 
likelihood~\eqref{eq:MF_likelihood}, and the independence assumptions, the 
joint evidence $p(\Y_{1:F})$ is given by~\eqref{eq:joint_evidence} where 
$\bSigma_{\y_{f}} = \bSigma_{\bbeta_{f}} + \A_{f}^{o} \bGamma \A_{f}^{o H}$.
Taking the logarithm and ignoring constant terms we have
\begin{align}
\log \,p(\Y_{1:F}) & \propto \sum_{f=1}^{F} \Big( - L \log |\bSigma_{\y_{f}}| - \text{Tr}(\Y_{f}^{H} \bSigma_{\y_{f}}^{-1} \Y_{f}) \Big).
\end{align}
To estimate $\hat{\bgamma}$ we maximize the joint evidence:
\begin{align}
\hat{\bgamma} &= \underset{\bgamma}{\arg \, \max} \, \log \, p(\Y_{1:F}) \\
&= \underset{\bgamma}{\arg \, \min} \Bigg\{ \sum_{f=1}^{F} L \log |\bSigma_{\y_{f}}| +
	\text{Tr}(\Y_{f}^{H} \bSigma_{\y_{f}}^{-1} \Y_{f}) \Bigg\}
\end{align}
To obtain a minimum, we apply the derivative results as before and equate the derivative 
of this objective function to zero giving the update rule
\begin{align}
& \frac{\partial}{\partial \gamma_{m}} \Bigg\{ \sum_{f=1}^{F} L \log |\bSigma_{\y_{f}}| +
	\text{trace}(\Y_{f}^{H} \bSigma_{\y_{f}}^{-1} \Y_{f}) \Bigg\} = 0 \\
\gamma_{m}^{\text{new}} & = \gamma_{m}^{\text{old}}  
	\Bigg( \frac{\sum_{f=1}^{F} \trace \Big( \bSigma_{\y_{f}}^{-1} [\bSigma_{f,m}^{e} + \a_{f,m}^{o} \a_{f,m}^{o H}] \bSigma_{\y_{f}}^{-1} \S_{\y_{f}} \Big)}
	{\sum_{f=1}^{F} \trace \Big(  \bSigma_{\y_{f}}^{-1} [\bSigma_{f,m}^{e} + \a_{f,m}^{o} \a_{f,m}^{o H}] \Big)} \Bigg)^{b}. \label{eq:gamma_update_MF}
\end{align}
In this multi-dictionary formulation, a unified update rule is obtained that combines 
all the observations together from different dictionaries. The single dictionary update 
rule~\eqref{eq:gamma_update} is obtained using $F = 1$.

\subsection{Special cases}
\label{sec:special_cases}

We consider special cases of \eqref{eq:linear_model_new} with $\x^{e} = \0$ and/or $\A^{e} = \0$:
\begin{itemize}
\item {\bf SBL}: when both $\x^{e} = \0$ and $\A^{e} = \0$ we get the regular SBL~\cite{tipping2001, wipf2007}, Eq \eqref{eq:linear_model}
\item {\bf SBL-A}: when only $\A^{e}$ is non-zero ($\x^{e} = \0$) gives
\begin{align}
\y &= \A^{o} \x^{o} + \A^{e} \x^{o} + \n
\label{eq:model_SBL_A}
\end{align}
signifying errors in the sensing matrix $\A$.
\item {\bf SBL-x}: when only $\x^{e}$ is non-zero ($\A^{e} = \0$) gives
\begin{align}
\y &= \A^{o} \x^{o} + \A^{o} \x^{e} + \n
\label{eq:model_SBL_x}
\end{align}
signifying errors in the weights $\x$.
\end{itemize}
Both SBL-A and SBL-x can be combined with the multi-dictionary SBL formulations 
SBL-MC and SBL-CC.

\subsection{Noise estimate}
Similar to $\gamma_{m}$, an update equation for $\sigma^{2}$ can be obtained 
using the derivative of the evidence with respect to $\sigma^{2}$. But this update 
is biased towards zero~\cite{wipf2007,liu2012,gerstoft2016}. Hence we use a 
stochastic maximum likelihood based method to estimate $\sigma^{2}$. 
Let $\A_{\cM}$ be formed by $K$ columns of $\A$ indexed by $\cM$, where the 
set $\cM$ indicates the location of non-zero entries of $\x$ with cardinality 
$|\cM| = K$. We can estimate $\cM$ using $\bgamma$ through thresholding or 
picking its highest entries. The noise variance estimate for $f$th dictionary is 
then~\cite{stoica1995, liu2012, gerstoft2016}
\begin{align}
\hat{\sigma}_{f}^{2} = \frac{1}{N-K} \trace \big( (\I_{N} - \A_{f,\cM} \A_{f,\cM}^{+}) \S_{\y_{f}} \big),
\label{eq:sigma_update}
\end{align}
where $\A_{\cM}^{+}$ denotes the Moore-Penrose pseudo-inverse. In~\cite{liu2012} 
a common noise estimate is used for all dictionaries (i.e. frequencies).

\subsection{Posterior}
Applying Bayes rule, the posterior for $\X$ is expressed as
\begin{align}
p(\X | \Y) &= \frac{p(\Y | \X) p(\X)}{p(\Y)}.
\end{align}
Since the prior is a Gaussian, the likelihood is approximated to be Gaussian, 
and the snapshots are independent, the posterior approximately is Gaussian 
with density given by
\begin{align}
p(\X | \Y) & \approx \prod_{l = 1}^{L}{\cC \cN (\x_{l}; \bmu_{l}, \bSigma_{\x})}, \\
\bmu_{l} &= \bGamma \A^{H} \bSigma_{\y}^{-1} \y_{l}, \quad \forall \, l = 1 \ldots L, \\
\bSigma_{\x} &= \bGamma - \bGamma \A^{H} \bSigma_{\y}^{-1} \A \bGamma.
\end{align}
The posterior mean $\bmu_{l}$ provides an estimate of the amplitude and phase
of the weight vector at the $l$th snapshot and also is sparse. The posterior
covariance matrix $\bSigma_{\x}$ provides an estimate of uncertainty in the weights.

\section{Simulations and experimental data}
\label{sec:simulations}

\subsection{SBL implementation}

\begin{algorithm}
\caption{Multi-dictionary SBL algorithm}
\begin{algorithmic}[1]
\State Parameters: $\epsilon = 10^{-6}, N_{t} = 3000, b = 1$
\State Input: $\S_{\y_{f}}, \A_{f}^{o} \, \forall f$, $\gamma_{m}^{e}, \bSigma_{m}^{e} \, \forall m$
\State Initialization: $\gamma_{m}^{\text{old}} = 1, \, \forall m, \quad \hat{\sigma}_{f}^{2} = 0.1, \, \forall f$
\State {\bf for} $i = 1$ to $N_{t}$
	\State $\quad$ Compute: $\bSigma_{\y_{f}} = \bSigma_{\bbeta_{f}} + \A^{o}_{f} \bGamma^{\text{old}} \A_{f}^{o H} \, \forall f$
	\State $\quad$ $\gamma_{m}^{\text{new}}$ update $\forall m$ using \eqref{eq:gamma_update_MF}
	\State $\quad$ $\hat{\sigma}_{f}^{2}$ estimate $\forall f$ using \eqref{eq:sigma_update}
	\State $\quad$ If $\frac{||\bgamma^{\text{new}} - \bgamma^{\text{old}}||_{1}}{||\bgamma^{\text{old}}||_{1}} < \epsilon$, {\bf break}
	\State $\quad$ $\bgamma^{\text{old}} = \bgamma^{\text{new}}$
\State {\bf end}
\State Output : $\bgamma^{\text{new}}$
\end{algorithmic}
\label{alg:SBL}
\end{algorithm}

This section discusses the algorithmic implementation of the SBL update rules 
developed in Sect.~\ref{sec:sbl}. A pseudocode of the SBL-CC algorithm is 
given in Algorithm~\ref{alg:SBL}. A similar algorithm can be obtained for SBL-MC
by replacing \eqref{eq:gamma_update_MF} with 
\eqref{eq:gamma_update_MF2a}-\eqref{eq:gamma_update_MF2b}. In either
case, the single dictionary algorithm is obtained by setting $F = 1$.

Parameters $\epsilon$ and $N_{t}$ determine the error convergence criteria and 
the maximum number of iterations, respectively. We choose the power exponent 
in the update rule \eqref{eq:gamma_update_MF} to be $b = 1$ as used 
in~\cite{tipping2001,wipf2007}.

The inputs to the algorithm are the sample covariance matrices $\S_{\y_{f}}$, the 
sensing matrices $\A_{f}^{o}$, and \emph{tuning} parameters $\gamma_{m}^{e}$ 
and $\bSigma_{m}^{e}$.
The parameters to estimate, $\gamma_{m}$ and $\sigma_{f}^{2}$, are initialized to 
constant non-zero values. The parameter $\gamma_{m}$ can be dictionary-dependent, 
see Sect.~\ref{sec:MF_SBL} SBL sum-MF, in which case there is an additional loop over 
all the dictionaries (not shown here). The $\gamma_{m}$ are updated using 
\eqref{eq:gamma_update_MF}. K peak locations are identified from 
$\bgamma^{\text{new}}$ to construct $\A_{\cM}$ and the dictionary-dependent noise 
estimate \eqref{eq:sigma_update}. Though we assume $K$ to be known for 
estimating $\hat{\sigma}^{2}$, this can be avoided by using model order identification 
methods~\cite{liu2012}. 

We use beamforming to demonstrate the benefits of the proposed SBL algorithms.
Sparsity of SBL is measured by $\bgamma$. Since the beamforming dictionary has 
high coherence among neighboring columns, we only consider local peaks. A local 
peak is defined as an element which is larger than its adjacent elements. Since 
$\bgamma$ corresponds to the source power, it is treated as the angular power spectrum.

We consider the special cases in Section~\ref{sec:special_cases}, SBL-A \eqref{eq:model_SBL_A}
and SBL-x \eqref{eq:model_SBL_x}. Additionally we assume
\begin{align}
\bSigma_{m}^{e} &= \phi^{e} \, \I_{N}, \quad \forall m \label{eq:phi_model} \\
\gamma_{m}^{e} &= \gamma^{e}, \, \quad \forall m. \label{eq:gamma_model}
\end{align}
This simplifies the number of free parameters and allows for a systematic study. The 
use of constants $\phi^{e}$ and $\gamma^{e}$ is justified when all the errors have 
similar statistics. Substituting \eqref{eq:phi_model} in \eqref{eq:full_model_covariance} 
with $\gamma^{e} = 0$, both the noise covariance $\sigma^{2} \I_{N}$ and the error 
covariance $\bSigma_{m}^{e}$ are diagonal. Hence it is difficult estimating both 
$\phi^{e}$ and $\sigma^{2}$. Whereas substituting \eqref{eq:gamma_model} in 
\eqref{eq:full_model_covariance} with $\phi^{e} = 0$ results in structurally different 
covariances and hence an estimate of $\gamma^{e}$ might be possible from data. 
In this paper we explore a range of tuning parameter values. 

Ideally the actual values of $\phi^{e}$ and $\gamma^{e}$ would depend on the application 
of interest. Since $\phi^{e}$ corresponds to the variance of the additive errors in the
dictionary, a good choice of $\phi^{e}$ could be obtained by studying the variability
of the underlying physical processes generating the dictionary. Since $\gamma^{e}$ 
is the variance of the errors in $\x$ (which is significant at low SNR), its value
can be tuned based on the SNR. Precaution should be taken to not choose relatively 
high values for $\phi^{e}$ and $\gamma^{e}$ as they tend to smooth out $\bgamma$ 
and could suppress weaker sources.

\subsection{Beamforming}
In beamforming, the observed signal model is a linear combination of plane waves.
Since the number of sources (arrival angles) is small, finely dividing the angle space 
results in a sparse $\x$ of complex amplitudes. SBL is used to recover these arrival 
angles.

For a narrow-band signal of wavelength $\lambda$ and uniform sensor array
separation $d$, the sensing matrix columns are
\begin{align}
\a_{m}^{o} &= [1, e^{j 2\pi \frac{d}{\lambda} \text{sin}(\theta_{m})},
	\ldots, e^{j 2\pi \frac{(N-1)d}{\lambda} \text{sin}(\theta_{m})}]^{T} \label{eq:beamforming_ao}
\end{align}
for $m = 1 \ldots M$, where $\theta_{m}$ is the $m$th discretized angle. The angle 
space $[-90,90]^{\circ}$ is discretized with $1^{\circ}$ separation giving $M = 181$. 
We model a $N = 20$ sensor array. The array SNR per snapshot is defined as
\begin{align}
\text{SNR} &= 10 \log \frac{\E\{||\a_{ws}^{o} x_{ws}||_{2}^{2}\}}{\E\{||\n||_{2}^{2}\}},
\label{eq:snr}
\end{align}
where the subscript $ws$ denotes weak source.
In this section we use a single frequency (a single dictionary) with sensor 
separation $d = \frac{\lambda}{2}$. $L = 30 (> N)$ snapshots are processed.

\subsubsection{Two source example}

Consider two sources present at angles $[0, 75]^{\circ}$ with powers $[22, 20]$ dB.
The magnitudes are assumed constant and their phases are random and distributed 
uniformly per snapshot.

\begin{figure}[b]
\centering
\includegraphics[width = 0.95\columnwidth]{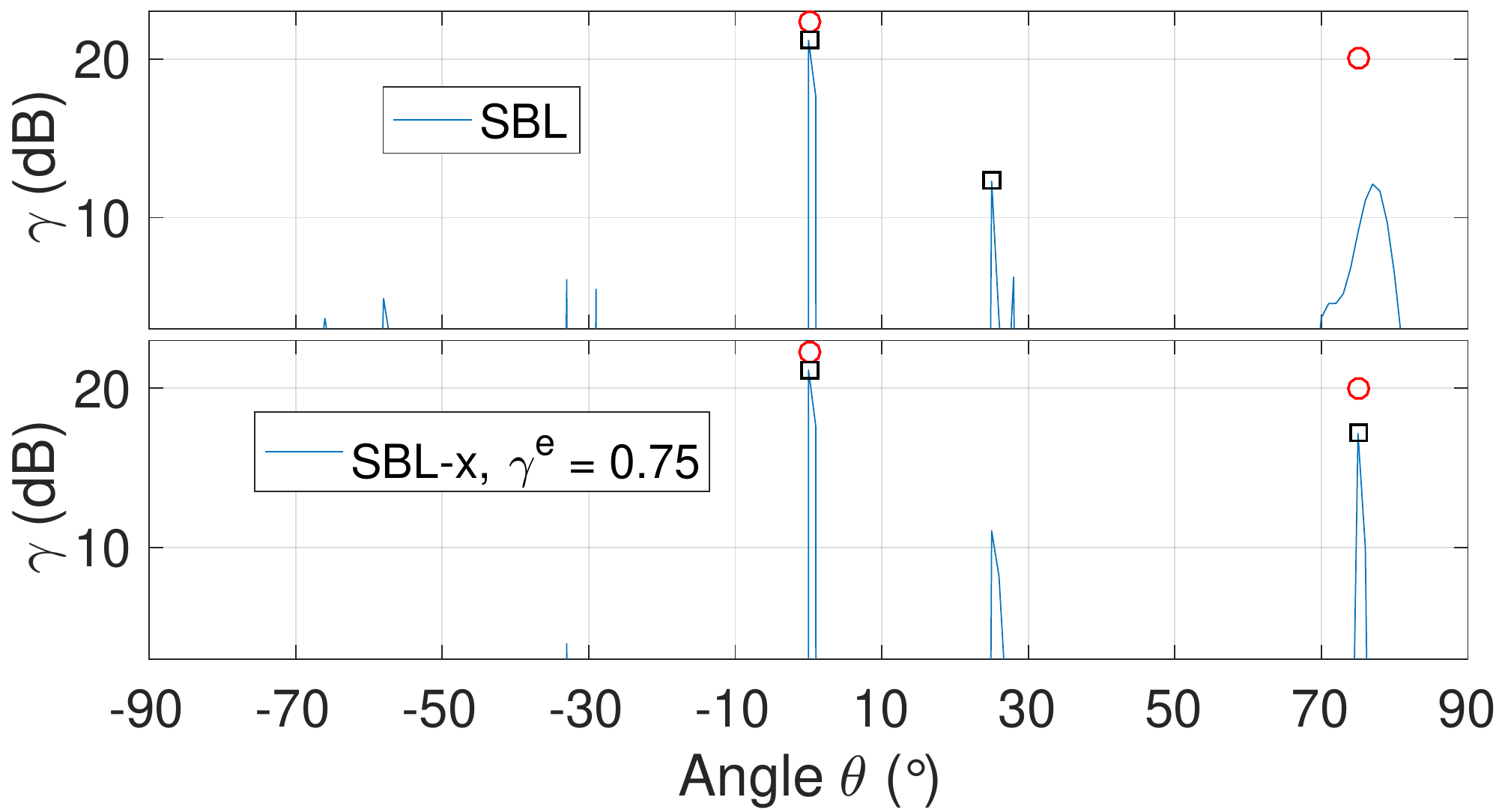}
\caption{Two-source example, SBL-x : Effect of parameter $\gamma^{e}$ on solution 
$\bgamma$ at SNR $3$ dB. True source location (red circles) and top two identified 
peaks (black squares) are indicated.}
\label{fig:2source_demo_snr_m6dB}
\end{figure}

Fig.~\ref{fig:2source_demo_snr_m6dB} shows $\bgamma$ for one run of the simulation
where SBL fails to correctly localize the peak at $75^{\circ}$ and changing the 
convergence parameters $\epsilon$ and $N_{t}$ in Algorithm~\ref{alg:SBL} does not 
change this. Due to high column coherence there is broadening of the peak at 
$75^{\circ}$ and hence redistribution of the peak energy. Using SBL-x 
($\gamma^{e} = 0.75$), the false peak is suppressed and the peak at $75^{\circ}$ is 
identified.

\begin{figure}[t]
\centering
\begin{subfigure}{0.99\columnwidth}
	\includegraphics[width = \textwidth]{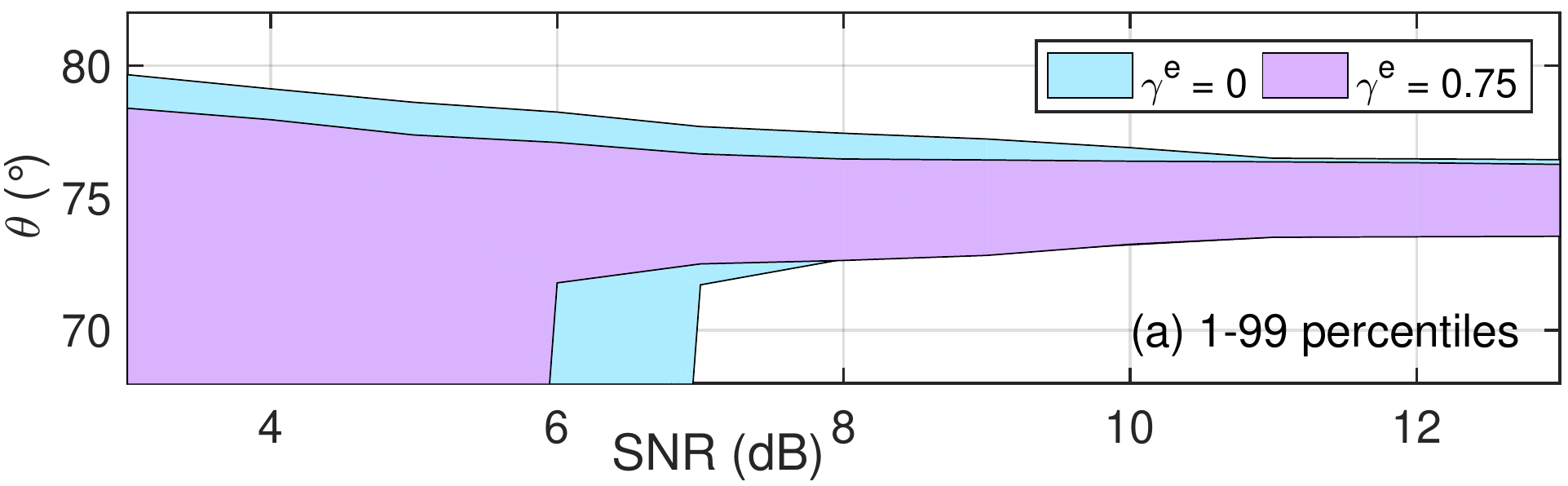}
\end{subfigure}
\begin{subfigure}{0.99\columnwidth}
	\includegraphics[width = \textwidth]{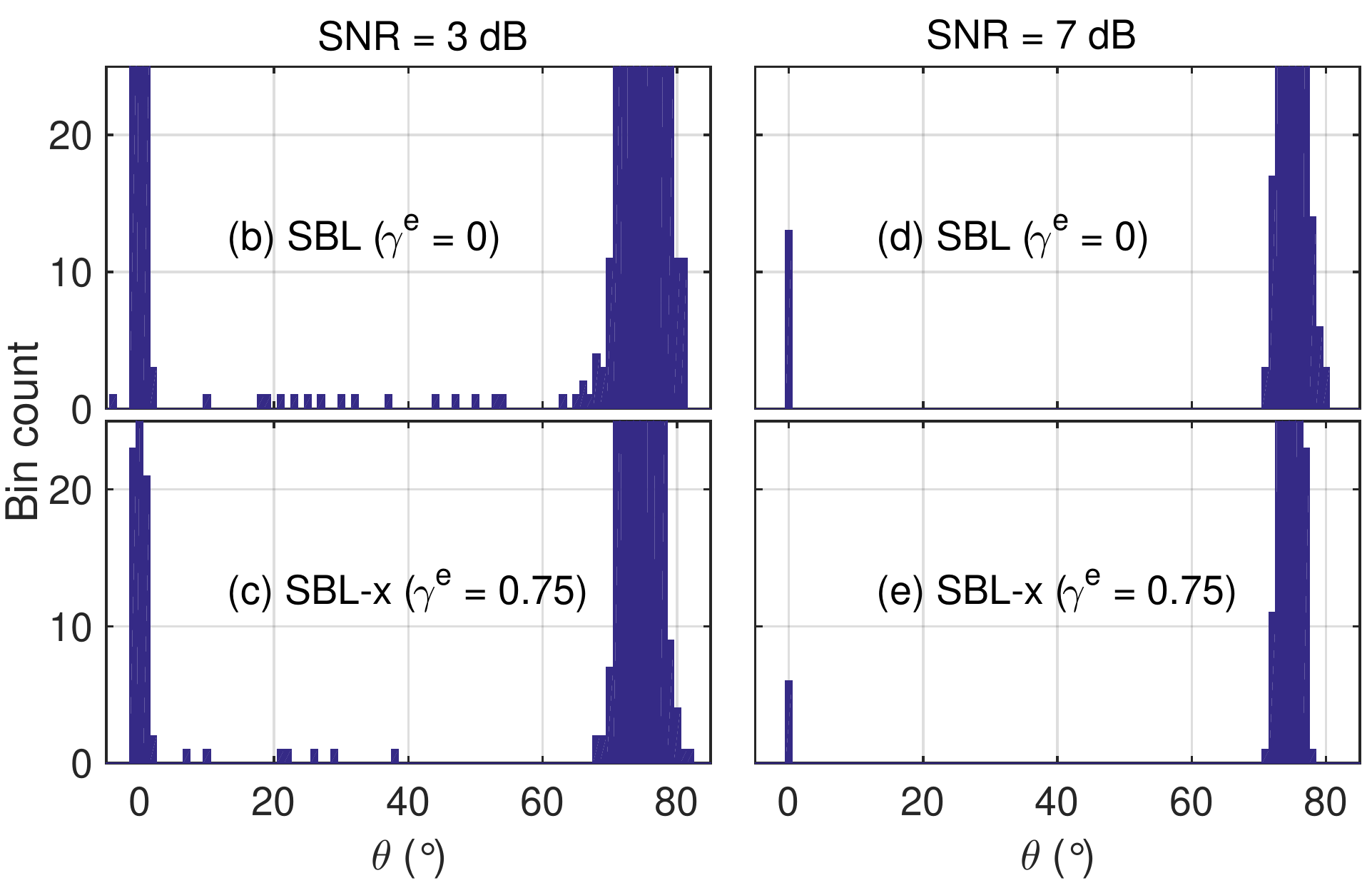}
\end{subfigure}
\caption{Two source example, SBL-x: (a) Shaded area shows the region between 1-99 
percentiles of the second strongest peak in $\bgamma$ over 2000 runs for $\gamma^{e} = 0, 0.75$.
Histograms of the second strongest peak location are shown for (b) $\gamma^{e} = 0$, (c) $\gamma^{e} = 0.75$ 
for SNR $3$ dB and for (d) $\gamma^{e} = 0$, (e) $\gamma^{e} = 0.75$ for SNR $7$ dB.}
\label{fig:2nd_source}
\end{figure}

These improvements in peak localization by SBL-x are illustrated using percentiles
of the second strongest peak location obtained from 2000 Monte Carlo runs in 
Fig.~\ref{fig:2nd_source}a. When $\gamma^{e} = 0.75$, the shaded area between 
the 1-99 percentiles shrinks, indicating better localization ability of SBL-x at low SNR. 
This reduction in the shaded area between the percentiles is due to fewer outlier 
points (one of these simulation runs was shown in Fig.~\ref{fig:2source_demo_snr_m6dB} 
where SBL-x is able to correctly localize the source at $75^{\circ}$ and avoid the 
outlier estimate at $25^{\circ}$). The localization improves with SNR as expected.
Histograms of the second strongest peak location for SNR $3$ dB and $7$ dB are 
shown in Fig.~\ref{fig:2nd_source}b, \ref{fig:2nd_source}c, \ref{fig:2nd_source}d and 
\ref{fig:2nd_source}e. Fewer outliers are observed for $\gamma^{e} = 0.75$ and the 
spread of the histogram is reduced around $75^{\circ}$ which is the true location of the
weaker source.

\subsubsection{Three source example}
\label{sec:three_source_example}

We consider three sources ($K=3$) located at angles $[-20, -15, 75]^{\circ}$ with 
powers $[10, 22, 20]$ dB. Following the model in Sect.~\ref{sec:prior_models}, the 
source amplitudes now are randomly sampled from a complex Gaussian with 
mean zero and variance equal to the source power. The weaker source ($10$ dB) 
close to the strongest source ($22$ dB) makes this challenging. In low SNR 
scenarios, this source can get masked by false peaks as seen in 
Fig.~\ref{fig:2source_demo_snr_m6dB}.

\begin{figure}[t]
\centering
\includegraphics[width = 0.98\columnwidth]{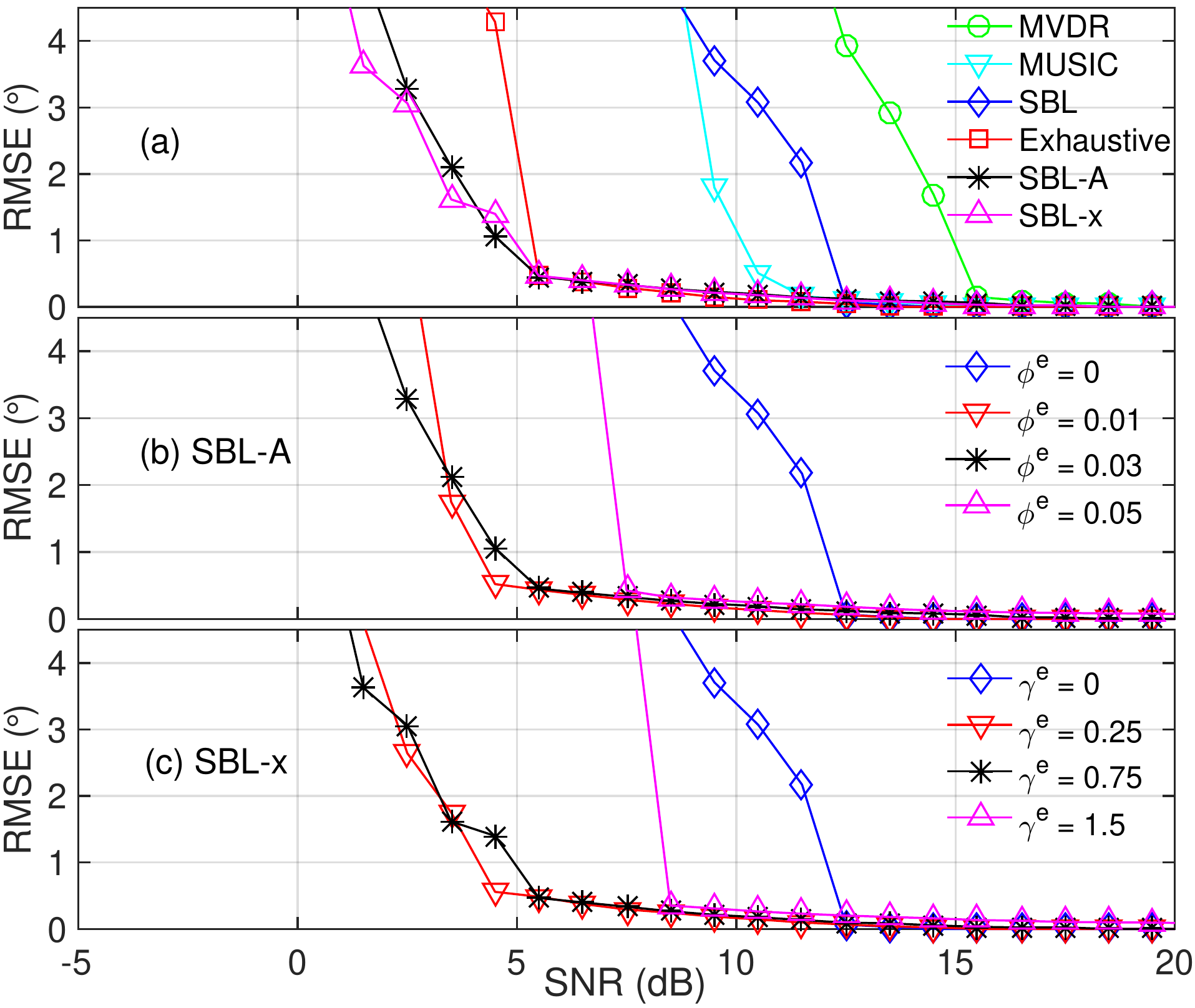}
\caption{Three source example: (a) RMSE comparison of DoA estimation methods 
MVDR, MUSIC, SBL, Exhaustive search, SBL-A, SBL-x using 2000 runs at 
each $\text{SNR}$ value. Effect of tuning parameters (b) $\phi^{e}$ and 
(c) $\gamma^{e}$ on DoA estimation for SBL-A and SBL-x respectively.}
\label{fig:SF_results_all}
\end{figure}

SBL is compared with traditional DoA estimation methods such as minimum
variance distortionless response (MVDR) and MUSIC in Fig.~\ref{fig:SF_results_all}a.
SBL outperforms MVDR while its performance is comparable to that of MUSIC. 
The root mean square error (RMSE) is
\begin{align}
\text{RMSE} &= \sqrt{\E (\hat{\theta}_{ws} - \theta_{ws})^{2}},
\end{align}
where $\theta_{ws}$ is the true and $\hat{\theta}_{ws}$ the estimated source 
angle of the weakest source. The expectation is computed from 2000 Monte Carlo 
runs. Since the weakest source likely fails first, it is appropriate restricting the 
RMSE metric to only this source. For traditional DoA methods, the estimated 
source angles ($\hat{\theta}_{k}$) are the top $3$ peaks in the angular power 
spectrum while, for SBL, they correspond to top $3$ peaks of $\bgamma$. 
The weakest of the top $3$ peaks is assigned to $\hat{\theta}_{ws}$.

Fig.~\ref{fig:SF_results_all}a compares the SBL-A and SBL-x algorithms with 
$\phi^{e} = 0.03$ and $\gamma^{e} = 0.75$. SBL-A shows reduced RMSE than 
SBL at low SNR indicating improved DOA estimation ability even though there 
is no perturbation in $\A$ (i.e. $\A^{e} = \0$). Also shown is the exhaustive 
search which finds the best DoA estimate $\cM_{0}$ by exhaustively solving 
the minimization problem
\begin{align}
\cM_{0} &= \underset{\cM}{\min} \; ||\Y - \A_{\cM} \tilde{\X}_{\cM}||_{\cal F} \label{eq:exhaustive_objective}
\end{align}
where $|| \cdot ||_{\cal F}$ is the Frobenius norm and 
$\tilde{\X}_{\cM} = \A_{\cM}^{+} \Y$. The objective function 
\eqref{eq:exhaustive_objective} is different from the SBL objective function
\eqref{eq:objective} and hence we expect different solutions. The SBL-A and SBL-x 
algorithms are able to outperform the exhaustive search method. An explanation 
for this is that at low SNR, the uncertainty models used by SBL-A and SBL-x better 
explain the noise in the solution allowing superior localization of the peaks.

The performance of the SBL-A and SBL-x algorithms for a range of $\phi^{e}$ and $\gamma^{e}$
are illustrated in Fig.~\ref{fig:SF_results_all}b and~\ref{fig:SF_results_all}c. SBL-A and SBL-x show
less sensitivity to the choice of parameters $\phi^{e} \in [0, 0.03]$ and $\gamma^{e} \in [0, 0.75]$
respectively. Further increasing $\phi^{e}$ and $\gamma^{e}$ the performance degrades as 
the model deviates significantly from the model generating the data.

\begin{figure}[t]
\centering
\includegraphics[width = 0.98\columnwidth]{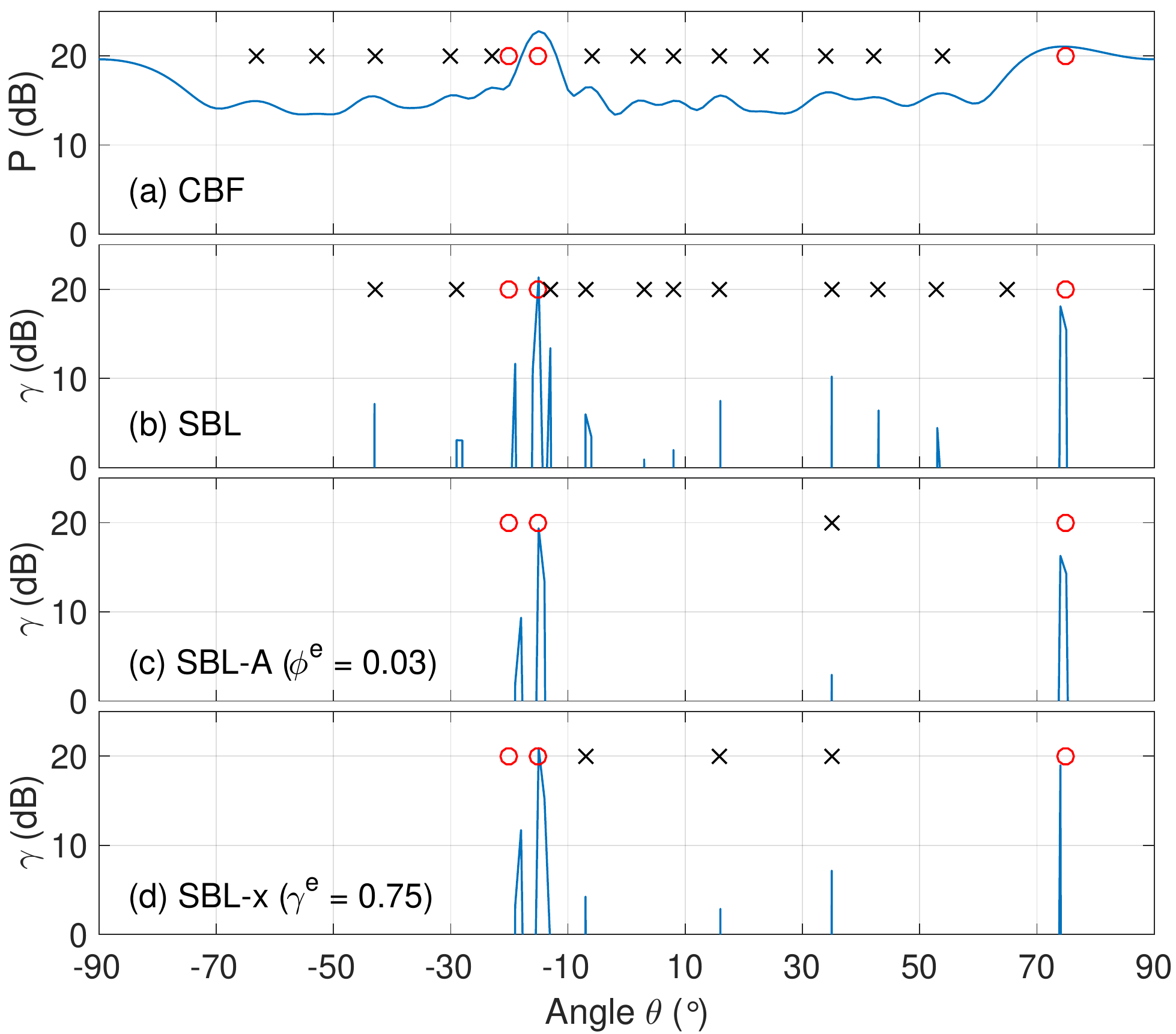}
\caption{Angular power spectrum of (a) CBF and $\bgamma$ of (b) SBL, (c) SBL-A, 
and (d) SBL-x when sources are located at $[-20,-15,75]^{\circ}$ with powers $[10,22,20]$ dB
and SNR $-2.5$ dB. True source locations (red circles) and false peaks (black crosses) 
are indicated.}
\label{fig:demo_figure}
\end{figure}

Fig.~\ref{fig:demo_figure} demonstrates the angular power spectrum for one run of the
simulation. For the conventional beamformer (CBF), the power spectrum is 
$\a_{m}^{o H} \S_{\y} \a_{m}^{o}$, and for SBL it is $\bgamma$. The CBF has broad peaks 
and the weaker peak at $-20^{\circ}$ is poorly identified. In regular SBL, many false peaks 
are present since the SNR is low. The strongest peak is split into two peaks in 
Fig.~\ref{fig:demo_figure}b. These false peaks compete with the weaker peak and 
represent errors in $\x$. SBL-A (Fig.~\ref{fig:demo_figure}c) and 
SBL-x (Fig.~\ref{fig:demo_figure}d) give improved performance and the false peaks are 
reduced.

\subsubsection{Mismatch analysis}

\begin{figure}[b]
\centering
\includegraphics[width = 0.95\columnwidth]{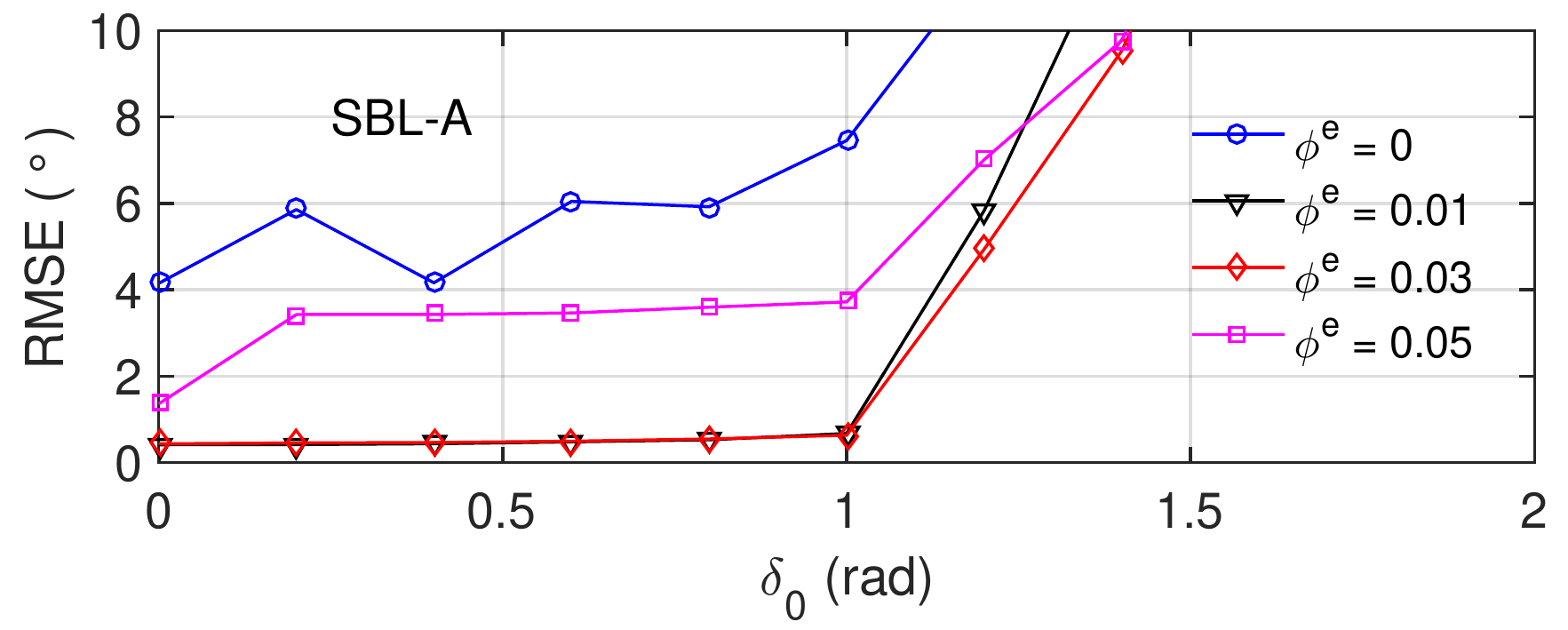}
\caption{Three source example, mismatch analysis: Effect of parameter $\phi^{e}$ 
in the SBL-A algorithm as the mismatch parameter $\delta_{0}$ is changed with 
SNR $5.5$ dB and 500 runs.}
\label{fig:results_chi2}
\end{figure}

SBL-A performance when the data is generated with mismatched dictionaries is studied 
by corrupting the dictionary with multiplicative noise, see Appendix A. The data is generated 
using multiplicative noise and processed using SBL-A which assumes additive Gaussian 
noise. Dictionaries are generated using the model
\begin{align}
\A &= \A^{o} \circ \A^{e}, \quad \A^{e}(m,n) = \exp (j \, \delta_{m}) \\
\delta_{m} & \sim \text{uniform} [-\frac{\delta_{0}}{2}, \frac{\delta_{0}}{2}].
\end{align}
The multiplicative noise parameter $\delta_{m}$ is the same for each column.
Each run of the simulation has a different $\A^{e}$.
The RMSE performance of SBL-A versus the parameter $\delta_{0}$ is in 
Fig.~\ref{fig:results_chi2}. Though the simulation scenario deviates from the 
modeling assumptions, SBL-A provides improvements.

\subsection{Aliasing suppression using multi-dictionary SBL}
\label{sec:aliasing_sbl}

SBL can be used to process multi-frequency spatial data in presence of aliasing. 
Each frequency has a different dictionary and the multi-dictionary analysis in 
Sect.~\ref{sec:MF_SBL} is used to process multi-frequency observations. 
Ref.~\cite{tang2011} discusses aliasing suppression for wideband signals 
using basis pursuit and orthogonal matching pursuit. We demonstrate aliasing 
suppression ability of SBL using both simulated and experimental data.

\subsubsection{Simulation analysis}

A large array aperture and hence a large sensor array spacing is desirable to 
obtain high resolution beamforming. A drawback of large array spacing is that 
it limits the highest frequency that can be processed without encountering 
aliasing. This drawback partially can be overcome by multi-dictionary SBL.

\begin{figure}[b]
\centering
\includegraphics[width = 0.98\columnwidth]{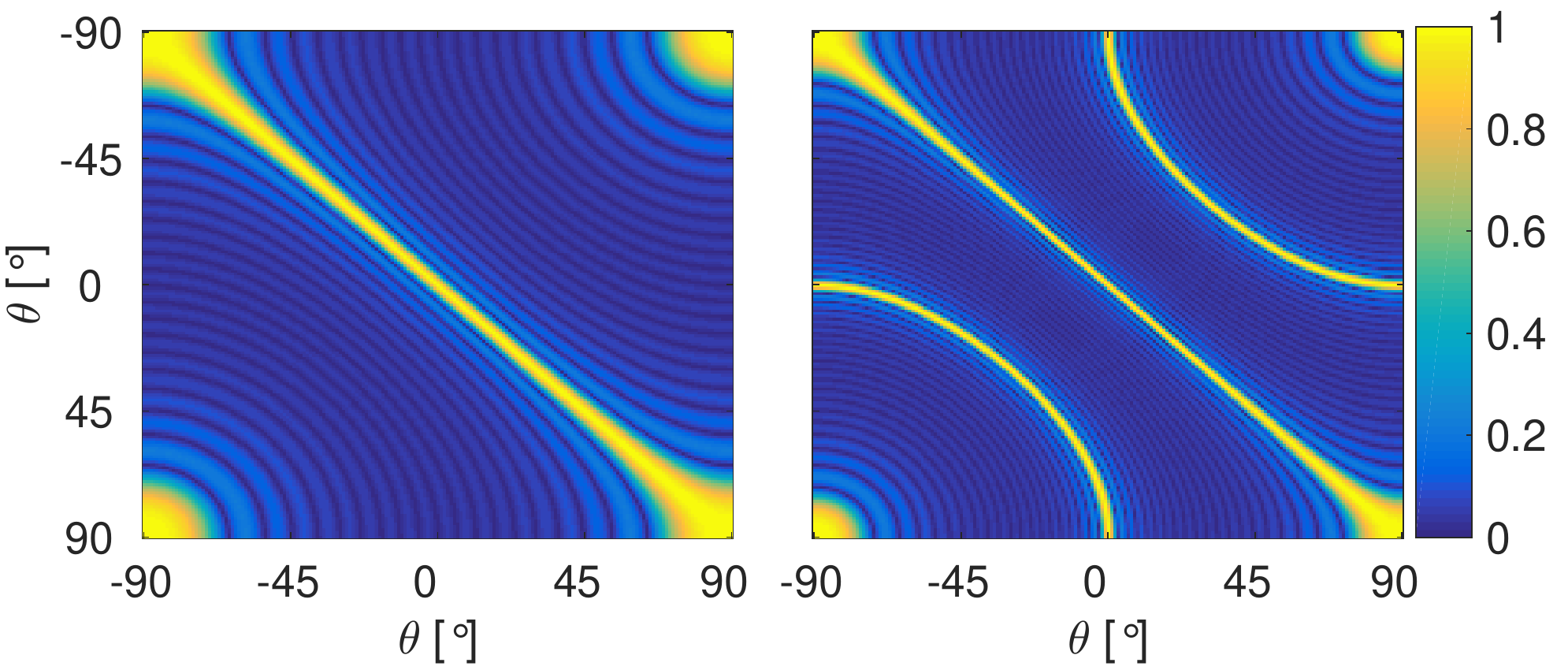}
\caption{Gram matrices for two array spacings : $d = \frac{\lambda}{2}$ (left), $d = \lambda$ (right).
Number of sensors $N = 20$.}
\label{fig:2F_gram_matrix}
\end{figure}

The Gram matrix $(\A^{H} \A)$ for two array spacings are shown in 
Fig.~\ref{fig:2F_gram_matrix}, $N = 20$. For a uniform linear array (ULA) 
spacing of $d = \frac{\lambda}{2}$ there is one main lobe for each angle. 
When the spacing is doubled, i.e. $d = \lambda$, grating (side) lobes 
appear which are a manifestation of aliasing.

\begin{figure}[t]
\centering
\begin{subfigure}{\columnwidth}
	\includegraphics[width = 0.99\textwidth]{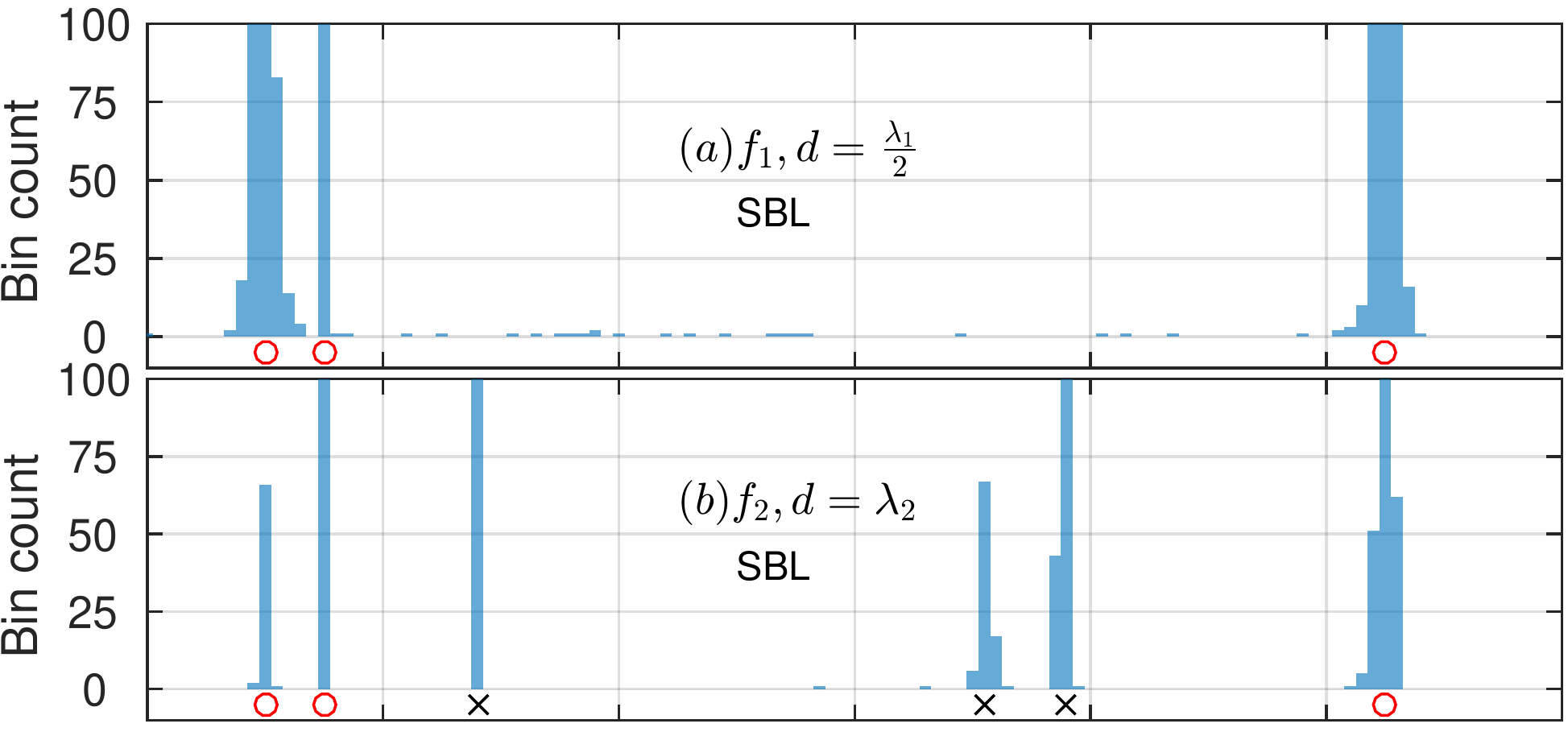}
\end{subfigure}
\begin{subfigure}{\columnwidth}
	\includegraphics[width = \textwidth]{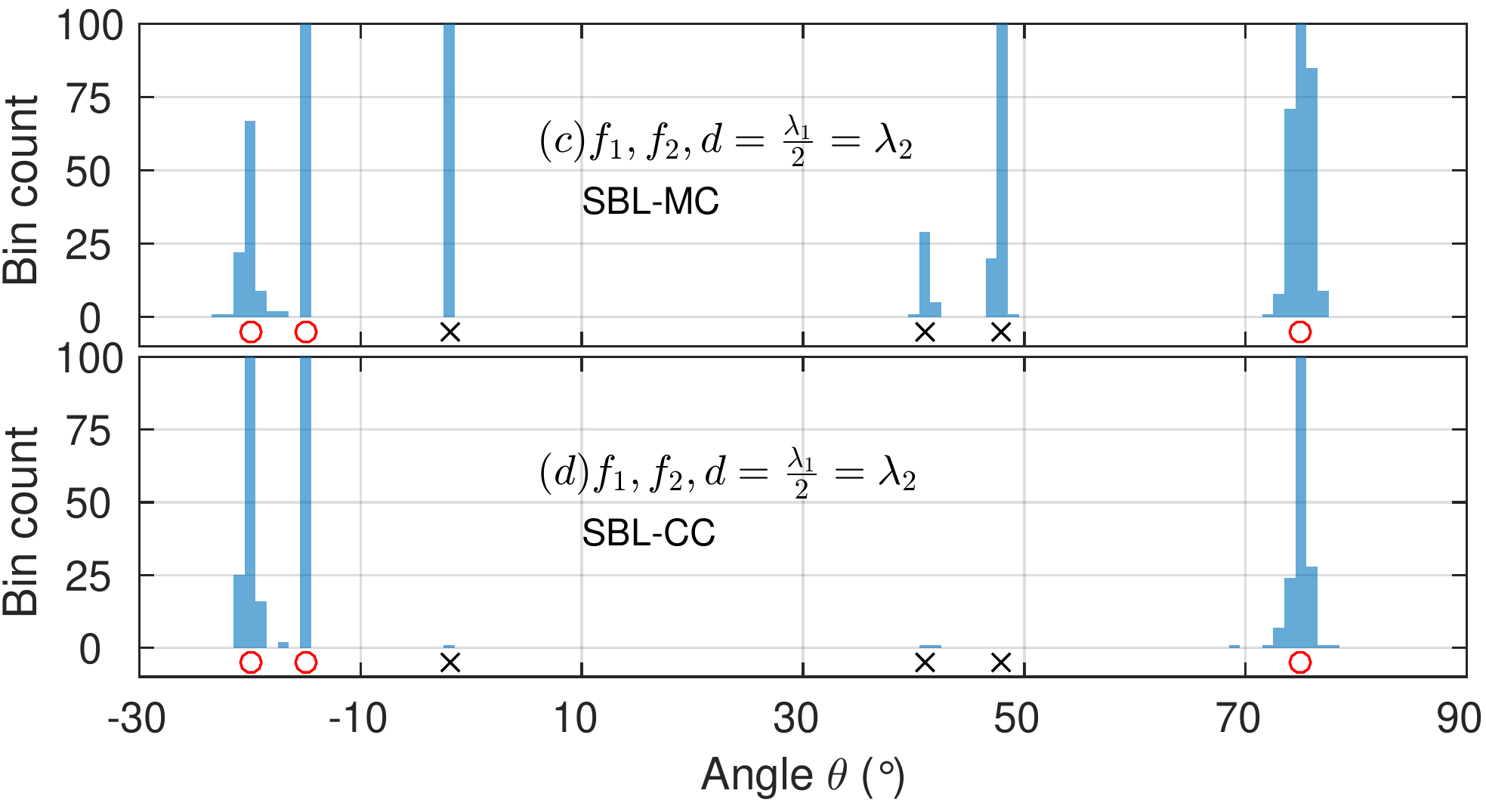}
\end{subfigure}
\caption{Aliasing analysis using histograms of the top three peaks: 
Single frequency (a) SBL, half-wavelength spacing (b) SBL, full-wavelength spacing.
Two frequencies (c) SBL-MC (d) SBL-CC. Number of sensors $N = 20$. 
Source (red circles) and aliased peak (black crosses) locations are indicated.}
\label{fig:aliasing_histogram}
\end{figure}

Consider the three source example in Sect.~\ref{sec:three_source_example}. 
Let $f_{1}$ and $f_{2} = 2 f_{1}$ be two frequencies with wavelengths 
$\lambda_{1}$ and $\lambda_{2} = \frac{\lambda_{1}}{2}$. The signal power 
is the same at each frequency for a given source. The histograms 
of the top three peaks obtained from $\bgamma$ are shown in 
Fig.~\ref{fig:aliasing_histogram} when observations from each frequency is 
processed independently using SBL. Aliasing is absent in 
Fig.~\ref{fig:aliasing_histogram}a since $d = \frac{\lambda_{1}}{2}$. 
Doubling the signal frequency with the same sensor spacing,
Fig.~\ref{fig:aliasing_histogram}b, gives aliased peaks. Higher frequency 
gives higher resolution but with additional aliased peaks. Thus SBL (and 
its variants SBL-A and SBL-x) cannot avoid aliasing when only a single 
frequency is used.

We now combine the observations from the two frequencies using multi-dictionary SBL 
when the sensor spacing is fixed at $d = \frac{\lambda_{1}}{2} = \lambda_{2}$. The two 
multi-dictionary SBL formulations are discussed in Sect.~\ref{sec:MF_SBL}. In SBL-MC,
observations from each frequency are processed independently and the multi-frequency
$\bgamma$ is obtained by summation~\eqref{eq:gamma_update_MF2b}. 
Fig.~\ref{fig:aliasing_histogram}c shows the histogram when SBL-MC is used. 
The bin count is significant at aliased locations and hence SBL-MC cannot suppress 
aliasing. The second multi-dictionary approach, SBL-CC, enforces a common sparsity 
profile by requiring $\bgamma$ to be the same across frequencies. The histogram obtained 
using SBL-CC is shown in Fig.~\ref{fig:aliasing_histogram}d. Since 
aliased peak locations are not shared across frequencies, they are suppressed by jointly 
processing multi-frequency observations using~\eqref{eq:gamma_update_MF}.

\subsubsection{Experimental data analysis}
\label{sec:experimental_data}

\begin{figure*}[t]
\centering
\begin{subfigure}{0.95\textwidth}
	\includegraphics[width = \textwidth]{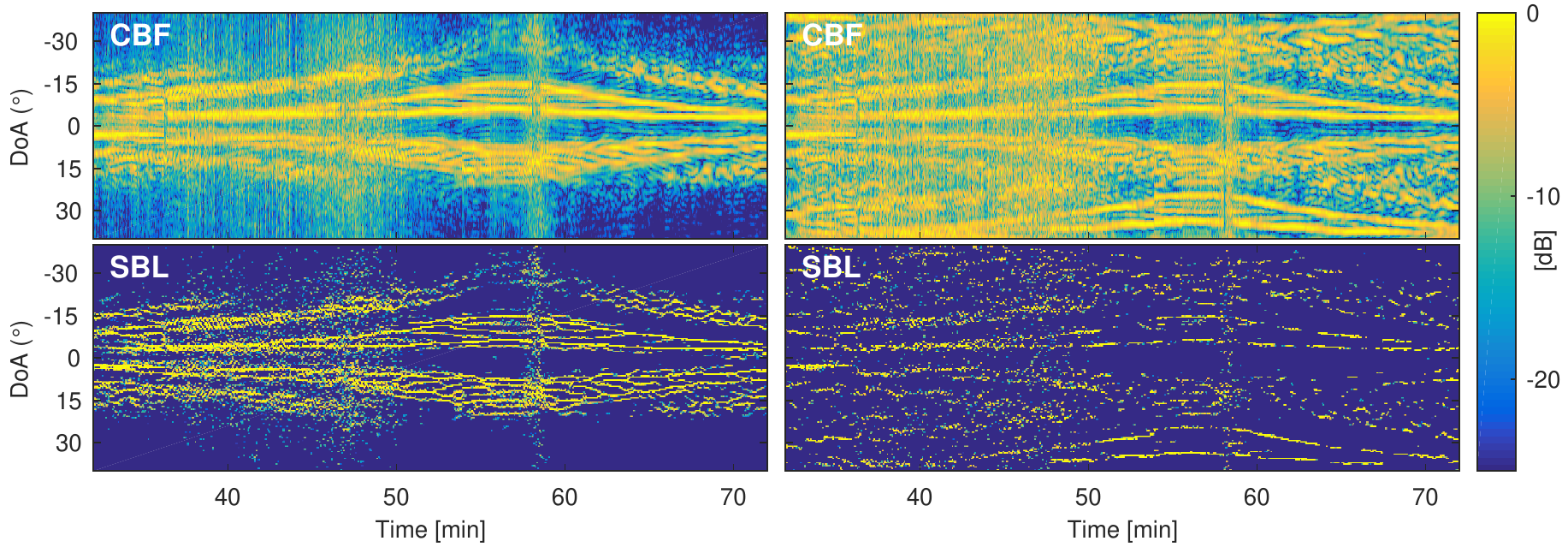}
	\caption{Single-frequency (388 Hz)}
	\label{fig:1F_Swellex_beamforming}
\end{subfigure}
\begin{subfigure}{0.95\textwidth}
	\includegraphics[width = \textwidth]{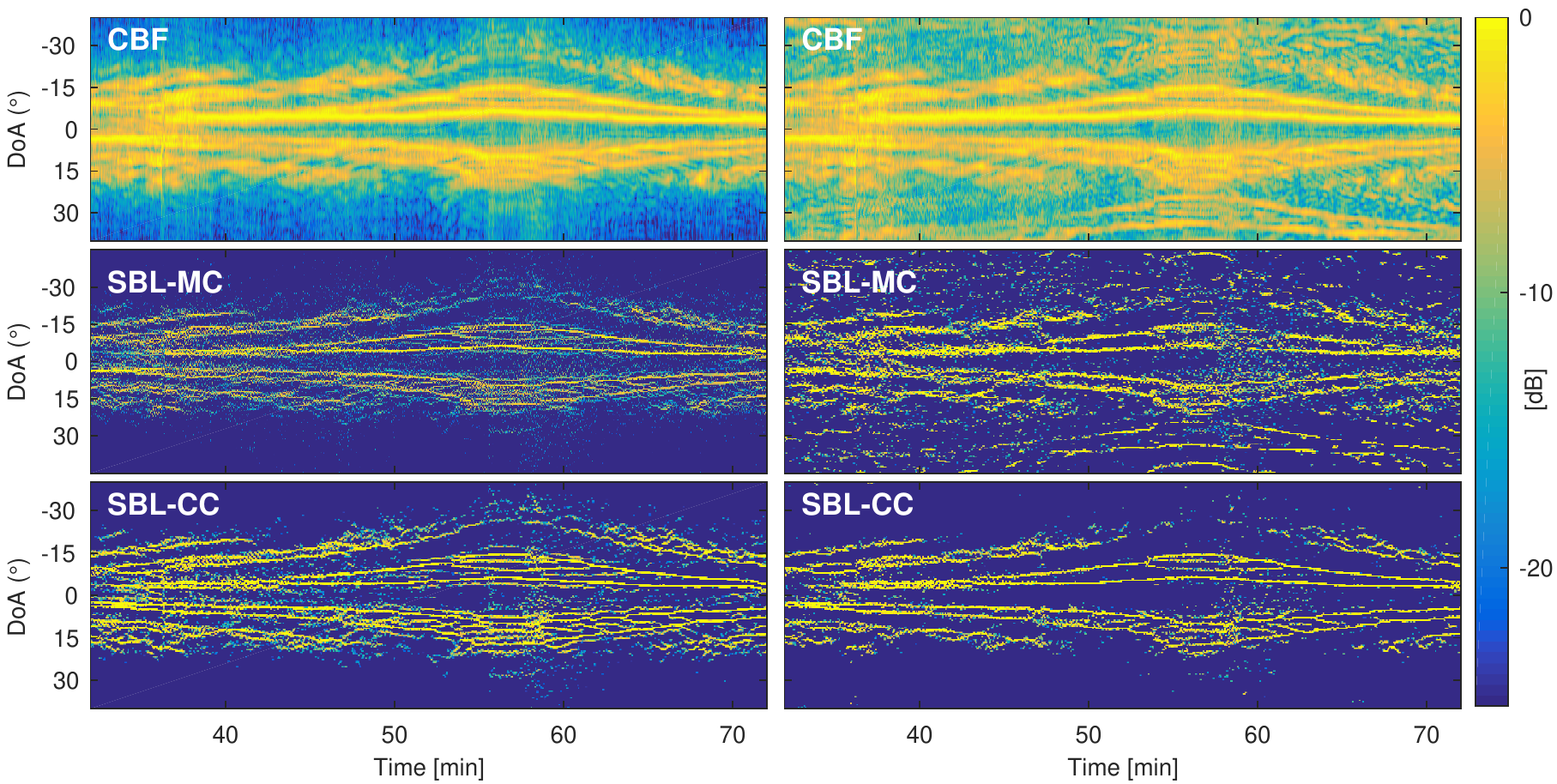}
	\caption{Multi-frequency (166, 283, and 388 Hz)}
	\label{fig:3F_Swellex_beamforming}
\end{subfigure}
\caption{(a) Single-frequency (388 Hz) and (b) multi-frequency (166, 283, and 388 Hz) 
analysis of SwellEx-96 Event S5 data using 63 (left column) and 21 (right column) elements 
of the array. In (a) the top row is CBF and bottom row is single frequency SBL. 
In (b) the top row is CBF, middle row is SBL-MC, and the bottom row is SBL-CC.
The columns of each of the panels are normalized.}
\end{figure*}

The high-resolution performance of SBL compared to CBF is validated with 
experimental data in a complex multi-path, shallow-water environment.  
The aliasing suppression ability of multi-dictionary SBL is demonstrated by 
processing a subset array.

The data is from the Shallow Water evaluation cell Experiment $1996$ (SWellEx-96) 
Event S5~\cite{gemba2016} collected on a $64$-element vertical line array. 
Element $43$ is excluded from processing. The array 
spans the lower part of the $212$ m watercolumn from $94$ to $212$ m with inter-sensor 
spacing $d = 1.875$ m. During the $77$ min Event S5, a deep source submerged at 
$60$ m was towed from $9$ km southwest to $3$ km northeast of the array at $5$ kn 
($2.5$ m/s).

The source was transmitting a set of ten frequencies with constant source levels 
of which the three frequencies $\{166, 283, 388\}$ Hz are used. The data are split 
into $2257$ overlapping segments, whereas a single segment is of 2.7 s duration. 
Snapshots are computed continuously from the data before being assigned to a 
segment. A FFT length of $2048$ samples (1.35 s) with $50$\% overlap results in 
$L=3$ snapshots for each segment with a FFT bin width of $0.75$ Hz. To 
accommodate Doppler shift, we search two adjacent FFT bins and extract the 
bin with maximum power.

Both the full array (64 elements, Array-1) and a subset (21 elements, Array-2) 
are used for processing. Array-2 is obtained by including every third element 
from Array-1 (Array-1 spacing $d$ and Array-2 spacing $3d$). By design, 
Array-1 suffers no aliasing whereas Array-2 suffers aliasing for frequencies 
above 133 Hz.

Single frequency (388 Hz) data is processed using both Array-1 and Array-2.
Fig.~\ref{fig:1F_Swellex_beamforming} shows CBF output power (top row) 
and $\bgamma$ for SBL (bottom row) as the source moves over time. Array-1 
processing does not suffer from aliasing (Fig.~\ref{fig:1F_Swellex_beamforming}, left)
and multi-path arrivals can be seen. SBL provides finer angular resolution than
CBF. Significant aliasing (Fig.~\ref{fig:1F_Swellex_beamforming}, right)
is present in both the SBL and CBF outputs when Array-2 is used. This 
aliasing is due to insufficient spacial sampling. Significant energy
is redistributed into aliased locations causing ambiguities in DoA estimation.

Combining three frequencies $\{166, 283, 388\}$ Hz and processing them 
from Array-1 and Array-2 is shown in Fig.~\ref{fig:3F_Swellex_beamforming}. 
Along with CBF output power (top row), the $\bgamma$ surfaces are shown for
SBL-MC (middle row) and SBL-CC (bottom row). Neither SBL nor CBF show 
any aliasing when Array-1 (Fig.~\ref{fig:3F_Swellex_beamforming}, left) data 
is processed. For Array-2 (Fig.~\ref{fig:3F_Swellex_beamforming}, right), CBF 
and SBL-MC both exhibit aliasing since the single frequency surfaces are averaged 
across frequencies. The relatively steep true arrivals around $\pm 20 ^\circ$ easily 
can get masked by the aliased arrivals causing DoA estimation errors. In comparison, 
SBL-CC shows no aliasing with Array-2 and the multi-path structure is preserved.
We note that in general there are slightly fewer peaks identified, when compared 
to the corresponding Array-1 results, because of the reduced array gain of Array-2.

\section{Conclusions}
\label{sec:conclusions}

The underdetermined system of linear equations in sparse processing is extended 
to account for errors in the sensing matrix and weights. The resulting non-Gaussian 
model was approximated as Gaussian to solve for the prior weight covariance using
SBL. An SBL update rule was developed which takes into account the statistics of 
uncertainty models. To estimate the noise variance a stochastic maximum likelihood 
based method was used.

We also developed SBL to process observations from multiple dictionaries when a 
portion of the support is common for all the weights. The first multi-dictionary SBL 
has dictionary-dependent priors which are summed to obtain a combined prior. The 
second multi-dictionary SBL requires the prior to be shared across dictionaries
giving a unified update rule.

Beamforming simulations for DoA estimation are used to demonstrate that false 
solutions can be removed at low SNR by explicitly accounting for errors in the 
sensing matrix and weights. Multi-frequency simulated and experimental data 
are processed using multi-dictionary SBL to recover DoAs in the presence of 
spatial aliasing. The multi-dictionary formulation with shared prior is able to 
avoid aliasing.

\section{Acknowledgement}
This work was supported by the Office of Naval Research Grant 
Nos. N00016-1-2341 and N00014-13-1-0632.

\section*{Appendix A - Multiplicative noise}
\label{sec:multiplicative_noise}
Perturbations in the sensing matrix can arise from multiplicative 
noise~\cite{paulraj1988, gershman1997, ringelstein2000}
\begin{align}
\A &= \A^{o} \circ \A^{e} \label{eq:A_mult_noise}
\end{align}
where $\A^{o}$ is a deterministic matrix, and $\A^{e}$ represents the multiplicative
error in $\A$. The notation $\circ$ denotes the Schur-Hadamard product of two 
matrices of same dimensions, i.e. the element-wise product of matrices given by
\begin{align}
[\A^{o} \circ \A^{e}]_{ij} &= [\A^{o}]_{ij} [\A^{e}]_{ij} 
\end{align}
A first order expansion of above multiplicative model \eqref{eq:A_mult_noise} is
\begin{align}
\A \approx \A^{o} \circ (\1 +  \A^{e_{1}}) &= \A^{o} + \A^{o} \circ \A^{e_{1}} \label{eq:taylor_exp1} \\
&= \A^{o} + \A^{e_{2}}. \label{eq:taylor_exp2}
\end{align}
where $\1$ denotes a matrix of all ones and $\A^{e_{2}} = \A^{o} \circ \A^{e_{1}}$. 
The model in~\eqref{eq:taylor_exp1} has been studied 
in~\cite{paulraj1988, gershman1997, ringelstein2000} and the model in~\eqref{eq:taylor_exp2} 
has been studied in~\cite{viberg1994a,viberg1994b,herman2010, zhu2011, teke2013, zhang2016}.

\bibliographystyle{IEEEtran}
\bibliography{sbl_bibtex}

\end{document}